\documentclass [12pt]{article}
\usepackage{graphicx,amssymb,amsmath,bm}
\usepackage[symbol*]{footmisc}
\usepackage[cp1251]{inputenc}
\usepackage[english]{babel}
\usepackage{booktabs}

\makeatletter
\def\@biblabel#1{#1.\hskip-0.3em}
\makeatother

\begin{document}
\def\refname{\normalsize \centering \mdseries \bf REFERENCES}
\def\abstractname{Abstract}

\begin{center}
{\large \bf Quasi-elastic scattering of $^{6}$\!He, $^{7}$\!Be, and $^{8}$\!B\\
nuclei by $^{12}$\!C nuclei}
\end{center}

\begin{center}
\bf\text{V.~I.~Kovalchuk}
\end{center}

\begin{center}
\small
\textit{Department of Physics, Taras Shevchenko National University, Kiev 01033, Ukraine}
\end{center}

\begin{abstract}
The observed cross sections of quasi-elastic scattering of $^{6}$\!He, $^{7}$\!Be, and $^{8}$\!B
nuclei by $^{12}$\!C nuclei are described within the framework of the diffraction nuclear model
and the model of nucleus-nucleus scattering in the high-energy approximation with a double folding
potential, for intermediate energies of the incident particles. The calculations make use of
realistic distributions of nucleon densities and take account of the Coulomb interaction and inelastic
scattering with excitation of low-lying collective states of the target.
\end{abstract}

\bigskip
\begin{center}
\bf{1.~INTRODUCTION}
\end{center}
\smallskip

In the last two decades the spectroscopy of light exotic nuclei near the nucleonic stability boundary
has been the subject of enhanced interest (see~\cite{1} and the references therein). One of the reasons
for such attention has been the discovery of a new property of the indicated nuclei -- the phenomenon
of a nuclear halo. Analysis of the first experiments on the interaction of neutron-rich isotopes of He,
Li, and Be with stable target nuclei~\cite{2} has led to the conclusion that a long-tail distribution
of nucleon density exists on such nuclei, arising from the small binding energy of the outer nucleons.
In experimental studies of the properties of such exotic nuclear systems, together with processes of
fragmentation, breakup, nucleon transfer, etc., an additional important source of information on the
density distribution of matter in halo nuclei is provided by elastic scattering. It should be noted that
the conditions of an experiment involving the participation of halo nuclei are often such that when
the particles are recorded a distinction is not made between the contributions of elastic and inelastic
processes. For this reason, such scattering can be called quasi-elastic.

For the purposes of a theoretical analysis of experiments on the scattering of exotic nuclei, most frequent
use is made of the Glauber formalism and the coupled-channels method. The present work compares the
possibilities of two approaches: the diffraction scattering model~\cite{3}, developed further in~\cite{4,5},
and the nucleus-nucleus scattering model in the high-energy approximation (HEA) with a double folding
potential~\cite{6}. As the object of analysis, we chose experiments on quasi-elastic scattering of
$^{6}$He~\cite{7}, $^{7}$Be, and $^{8}$B nuclei by $^{12}$C nuclei~\cite{8}. The experimental data
from these works contain an admixture of the inelastic channel due to excitation of collective states of
the target $2^{+}$ (4.44~MeV) and $3^{-}$ (9.64~MeV).

All of the calculations that follow were made in the center-of-mass (c.m.s.) system using the system of
units \mbox{$\hbar=c=1$}. Particle spin was not taken into account.

\bigskip
\begin{center}
\bf{2.~DIFFRACTION MODEL OF SCATTERING OF WEAKLY BOUND TWO-CLUSTER NUCLEI BY NUCLEI}
\end{center}

\begin{center}
\bf{A.~Elastic scattering}
\end{center}
\smallskip

The weakly bound nuclei $^{6}$He, $^{7}$Be, and $^{8}$B can be described as two-cluster systems
($^{4}\text{He}+2n$, $^{4}\text{He}+^{3}\text{He}$, and $^{7}\text{Be}+p$, respectively).
The scattering amplitude of such nuclei in the diffraction approximation is the sum of the amplitudes
of single and double scattering of clusters
\begin{equation}
G(\mathbf{q})=G^{(1)}(\mathbf{q})+G^{(2)}(\mathbf{q}),
\label{eq1}
\end{equation}
where $\mathbf{q}$ is the momentum transfer. The first term on the right-hand side of Eq.~(\ref{eq1})
has the form~\cite{4,5}
\begin{equation}
G^{(1)}(\mathbf{q})=ik\{\Phi_{0}(\beta_{2}\mathbf{q})u_{1}(\mathbf{q})+
\Phi_{0}(\beta_{1}\mathbf{q})u_{2}(\mathbf{q})\},
\label{eq2}
\end{equation}
where $k$ is the momentum of the incident nucleus, $\Phi_{0}$ is its structure form factor,
$\beta_{1,2}=m_{1,2}/(m_{1}+m_{2})$, and $m_{j}$ is the mass of the $j\text{th}$ cluster $(j=1,2)$.
The function $u_{j}(\mathbf{q})$ in expression Eq.~(\ref{eq2}) is the sum of the nuclear
contribution $u_{j}^{(N)}(\mathbf{q})$ and the Coulomb contribution $u_{j}^{(C)}(\mathbf{q})$
to the single-scattering amplitude for scattering of the $j\text{th}$ cluster
\begin{equation}
u_{j}^{(N)}(\mathbf{q})=\frac{1}{2\pi}\int d^{(2)}\mathbf{s}_{j} \exp(i\mathbf{q}\mathbf{s}_{j})
\omega_{j}(s_{j}),\quad
u_{j}^{(C)}(\mathbf{q})\equiv u_{j}^{(C)}(q)=2in_{j}g_{j}(qR_{j})/q^{2},
\label{eq3}
\end{equation}
where $\mathbf{s}_{j}$ is the impact parameter, $\omega_{j}(s_{j})$ is the profile function,
\mbox{$R_{j}=r_{0}(A_{t}^{1/3}+A_{j}^{1/3})$} is the radius of the cluster-nucleus interaction,
$A_{t}(A_{j})$ is  the mass number of the target nucleus (the $j\text{th}$ cluster), $n_{j}$ is
the corresponding Sommerfeld parameter, and
\begin{equation}
g_{j}(x)=\frac{\Gamma(1+in_{j})}{\Gamma(1-in_{j})}(2/x)^{2in_{j}}-
x\int_{0}^{1}d\xi\,\xi^{2in_{j}}J_{1}(x\xi).
\label{eq4}
\end{equation}

The second term in expression (\ref{eq1}) is a sum of double-scattering amplitudes
\begin{equation}
G^{(2)}(\mathbf{q})=-\frac{ik}{2\pi}\Big(u_{12}^{(NN)}(\mathbf{q})+
u_{12}^{(CN)}(\mathbf{q})+u_{12}^{(NC)}(\mathbf{q})\Big).
\label{eq5}
\end{equation}
Each of the functions $u_{12}$ entering into expression (5), is the contribution to
$G^{(2)}(\mathbf{q})$ of double scattering of the indicated pair of clusters 12, where the
superscripts on the functions $u_{12}$ indicate the type of interaction through which the
contribution of the given pair is realized ($N$ is nuclear, $C$ is Coulomb):
\begin{equation}
u_{12}^{(NN)}(\mathbf{q})=\int d^{(2)}\mathbf{p}\,\Phi_{0}(\mathbf{p}-\beta_{1}\mathbf{q})
u_{1}(\mathbf{p})u_{2}(\mathbf{p}-\mathbf{q}),
\label{eq6}
\end{equation}
\begin{equation}
u_{12}^{(CN)}(\mathbf{q})=\frac{n_{1}}{\sqrt{2\pi\lambda}}\int_{0}^{\infty}dp
\int_{0}^{\pi}d\vartheta \sin{\vartheta}\,\Phi_{0}(\mathbf{p}-\beta_{1}\mathbf{q})
\Lambda{(p,\vartheta,\lambda)} u_{2}^{(N)}(\mathbf{p}-\mathbf{q}) g_{1}(pR_{1}),
\label{eq7}
\end{equation}
and $u_{12}^{(NC)}$ is obtained from $u_{12}^{(CN)}$ by making the replacements \mbox{$n_{1}\to n_{2}$},
\mbox{$u_{2}^{(N)}\to u_{1}^{(N)}$}, and \mbox{$R_{1}\to R_{2}$} in expression (7). The function
\mbox{$\Lambda{(p,\vartheta,\lambda)}$} is given by the expression~\cite{4}
\begin{equation}
\Lambda{(p,\vartheta,\lambda)}=\exp\Big(\!-\frac{p^{2}\sin^{2}\vartheta}{16\lambda}\Big)
I_{0}\Big(\frac{p^{2}\sin^{2}\vartheta}{16\lambda}\Big),
\label{eq8}
\end{equation}
where $\lambda=3/(4 R_{\text{rms}}^{2})$ and $R_{\text{rms}}$ is the root-mean-square mass
radius of the incident nucleus.

\bigskip
\begin{center}
\bf{B.~Inelastic diffraction scattering with excitation of \\low-lying collective states of the target}
\end{center}
\smallskip

In~\cite{5} it was shown that the inelastic scattering amplitude (ISA) with excitation of low-lying
vibrational states of even-even nuclei $|IM\rangle$ having spin and its projection $|00\rangle$ in
their ground state can be represented in the form
\begin{equation}
F^{IM}(q)=f_{1}^{IM}(q)+f_{2}^{IM}(q),
\label{eq9}
\end{equation}
where $f_{j}^{IM}$ is the cluster-nucleus inelastic scattering amplitude (ISA):
\begin{equation}
f_{j}^{IM}=\exp{[2i\eta_{j}(R_{j})]}u_{j}^{IM}.
\label{eq10}
\end{equation}
Here $\eta_{j}(R_{j})=2n_{j}\ln{(k_{j}R_{j})}$ is the Coulomb phase ($k_{j}$ is the momentum of
the $j\text{th}$ cluster),
\begin{equation}
u_{j}^{IM}=\frac{ik_{j}}{2\pi}\int_{0}^{2\pi}d\phi\int_{R_{j}}^{R_{j}(1+Z_{IM}(\phi))}d{s_{j}}\,
s_{j}\exp{\![iqs_{j}\cos{\phi}]}\,\omega_{j}(s_{j}),
\label{eq11}
\end{equation}
\begin{equation}
Z_{IM}(\phi)=\displaystyle\frac{\beta_{I}\cos{M\phi}}{\sqrt{2I+1}}
   \left\{{\begin{array}{*{20}c}
   {i^{I+M}\sqrt{\displaystyle\frac{2I+1}{4\pi}}\displaystyle\frac{\sqrt{(I-M)!(I+M)!}}{(I-M)!!(I+M)!!}
    \quad \text{for}\,\,(I+M)\,\,\text{even},}  \\
   {0}\qquad\qquad\qquad\qquad\qquad\qquad\qquad\quad\,\, \text{for}\,\,(I+M)\,\,\text{odd}.  \\
\end{array}} \right.
\label{eq12}
\end{equation}
The quantity $\beta_{I}$ in expression~(\ref{eq12}) is the target deformation parameter. In the calculations
these quantities were set equal to $\beta_{2}=0.582$~\cite{9} and $\beta_{3}=0.365$~\cite{10}.

\bigskip
\begin{center}
\bf{3.~NUCLEUS-NUCLEUS SCATTERING IN THE HIGH-ENERGY APPROXIMATION WITH A DOUBLE FOLDING POTENTIAL}
\end{center}

The nucleus-nucleus scattering amplitude at energies of 10-100 MeV/nucleon without taking the Coulomb
interaction into account has the form
\begin{equation}
f(q)=ik\int db\,b\,J_{0}(qb)\,\omega{(b)},
\label{eq13}
\end{equation}
where the profile function $\omega{(b)}$ is expressed in terms of the eikonal phase $\chi{(b)}$:
\begin{equation}
\omega{(b)}=1-\exp{[i\chi{(b)}]},\quad
\chi{(b)}=-\frac{1}{v}\int_{-\infty}^{\infty}dz\,U(\sqrt{b^2+z^2}).
\label{eq14}
\end{equation}
Here $v$ is the velocity of relative motion of the nuclei, $U$ is the optical nucleus-nucleus
semi-microscopic potential
\begin{equation}
U(r)=N_{V}V^{DF}(r)+iN_{W}W(r),
\label{eq15}
\end{equation}
where $N_{V}$ and $N_{W}$ are normalization factors, $V^{DF}(r)$ is the double folding potential,
and $W(r)$ is the imaginary part of the optical potential, which is modeled by a Woods-Saxon dependence
with three free parameters: the depth $W_0$, radius $R_{W}=r_{W}(A_{p}^{1/3} + A_{t}^{1/3})$,
and diffusivity $a_{W}$. The potential $V^{DF}$ is the sum of a direct part $(V^{D})$ and an exchange
part $(V^{EX})$~\cite{6}
\begin{equation}
V^{DF}(r)=V^{D}(r)+V^{EX}(r),
\label{eq16}
\end{equation}
which have the following form:
\begin{equation}
V^{D}(r)=\int d^{(3)}\mathbf{r}_{p}d^{(3)} \mathbf{r}_{t}\rho_{p}{(\mathbf{r}_{p})}
\rho_{t}{(\mathbf{r}_{t})}\nu_{NN}^{D}(s),\quad
\mathbf{s}=\mathbf{r}+\mathbf{r}_{t}-\mathbf{r}_{p},
\label{eq17}
\end{equation}
\begin{equation}
V^{EX}(r)=\int d^{(3)}\mathbf{r}_{p}d^{(3)} \mathbf{r}_{t}
\rho_{p}{(\mathbf{r}_{p},\mathbf{r}_{p}+\mathbf{s})}
\rho_{t}{(\mathbf{r}_{t},\mathbf{r}_{t}-\mathbf{s})}\nu_{NN}^{EX}(s)
\exp{[i\mathbf{K}(r)\mathbf{s}/M]}.
\label{eq18}
\end{equation}
Here $\rho_{p,t}{(\mathbf{r}_{p,t})}$ is the single-particle density (local in Eq.~(\ref{eq17})
or non-local in Eq.~(\ref{eq18}), see~\cite{6}) of the incident nucleus $(p)$ and the target nucleus $(t)$,
$\nu_{NN}$ is the nucleon-nucleon effective potential,
$K(r)=\Big[2Mm\Big(E-V^{DF}(r)-V_{C}(r)\Big)\Big]^{1/2}$ is the local nucleus-nucleus momentum, where
$M=A_{p}A_{t}/(A_{p}+A_{t})$, $m$ is the mass of a nucleon, $E$ is the kinetic energy of the projectile
(in the c.m.s.), and $V_{C}$ is the Coulomb potential.

The scattering amplitude taking the Coulomb interaction into account can be represented as
\begin{equation}
F(q)=f(q)+f_{C}(q),
\label{eq19}
\end{equation}
where $f(q)$ is amplitude~(\ref{eq13}), and $f_{C}(q)$ has the form~\cite{3}
\begin{equation}
f_{C}(q)=-\frac{2nk}{q^2}
\Big[
\frac{\Gamma(1+in)}{\Gamma(1-in)}\Big(\frac{2}{qR}\Big)^{2in}-
qR\int_{0}^{1}d\xi\,\xi^{2in}J_{1}(qR\xi)
\Big].
\label{eq20}
\end{equation}
Here $n$ is the Sommerfeld parameter of the colliding nuclei, $k$ is their relative momentum,
and $R$ is a quantity defined above.

\bigskip
\begin{center}
\bf{4.~RESULTS OF CALCULATIONS AND THEIR ANALYSIS. CONCLUSIONS}
\end{center}

In the construction the profile functions in formula~(\ref{eq3}) we used an approach analogous to that
described in~\cite{11}, namely, we joined a tail to the unit step function, this tail consisting of the
nucleon density distribution of the corresponding cluster, normalized to unity, $\rho_{j}^{N}(x)$:
\begin{equation}
\omega_{j}(s_{j})=(\delta_{j}-i\gamma_{j})
\Big[
\Theta(R_{j}-s_{j})+\rho_{j}^{N}(s_{j}+R_{j})\Theta(s_{j}+R_{j})
\Big],
\label{eq21}
\end{equation}
where $\delta_{j}$ and $\gamma_{j}$ are the absorption and refraction parameters, respectively, and
$\Theta(x)$ is the Heaviside unit step function. The distributions $\rho_{j}^{N}(x)$ for the $^{6}$\!He
nucleus were constructed on the basis of data taken from~\cite{12,13} (LSSM is the large-scale shell model)
for the $^{7}$\!Be, and for the $^{8}$\!B nuclei, from~\cite{14}, where the nucleon density was calculated
by the density functional method.

In the calculations of double folding potential~(\ref{eq16}) for $\nu_{NN}$ we used the Paris
nucleon-nucleon potential in the CDM3Y6 form
\begin{equation}
\nu_{NN}(E,\rho,s)=g(E)F(\rho)\sum_{j=1}^{3}N_{j}\frac{\exp{(-\mu_{j}s})}{\mu_{j}s},
\label{eq22}
\end{equation}
\begin{equation*}
g(E)=1-0.003E/A_{p}\,,\quad
F(\rho)=C(1+\alpha\exp{(-\beta\rho)}-\gamma\rho),\quad
\rho=\rho_{p}+\rho_{t},
\end{equation*}
\begin{equation}
C=0.2658,\quad
\alpha=3.8033,\quad
\gamma=4.
\label{eq23}
\end{equation}

The potential parameters $N_{j}$ and $\mu_{j}$ are given in~\cite{15}. In the solution of the
nonlinear problem of finding the double folding potential, we used the algorithm and
computer code from~\cite{6}. Here we chose the nucleon density distribution of the $^{12}$\!C
target in the form of a symmetrized Fermi distribution~\cite{11}, and for the projectile we chose
the distribution to be the one calculated within the framework of the LSSM model ($^{6}$\!He) and
by the density functional method ($^{7}$\!Be and $^{8}$\!B).

Figure~1 plots the results of calculation of the ratios of cross sections $\sigma/\sigma_{R}$
($\sigma_{R}=(2kn)^2/q^4$ is the Rutherford cross section), where

(i) the dotted curve, $\sigma\equiv\sigma_{2}(q)$, is the contribution of inelastic scattering with
excitation of the $2^{+}$ level of the target:
\begin{equation}
\sigma_{2}(q)=\sum_{M=-I}^{I}|F^{IM}(q)|^{2},\quad
I=2;
\label{eq24}
\end{equation}

(ii) the dash-dot curve, $\sigma\equiv\sigma_{3}(q)$, is the contribution of inelastic
scattering with excitation of the $3^{-}$ level of the target:
\begin{equation}
\sigma_{3}(q)=\sum_{M=-I}^{I}|F^{IM}(q)|^{2},\quad
I=3;
\label{eq25}
\end{equation}

(iii) the thin continuous curve, $\sigma\equiv\sigma_{\text{el}}(q)$, is the elastic scattering
cross section:
\setlength\parindent{7ex}

$\sigma_{\text{el}}(q)=|G(\mathbf{q})|^{2}$ (Figs.~1$a$,~$c$, and~$e$ correspond to the
diffraction model),

$\sigma_{\text{el}}(q)=|F(q)|^{2}$ (Figs.~1$b$,~$d$, and~$f$ correspond to the HEA model);
\setlength\parindent{4ex}

\vspace{5mm}
(iv) the thick continuous curve, $\sigma\equiv\sigma_{\Sigma}(q)$, is the incoherent sum of
elastic scattering and inelastic scattering:
\setlength\parindent{7ex}

$\sigma_{\Sigma}(q)=|G(\mathbf{q})|^{2}+\sigma_{2}(q)+\sigma_{3}(q)$ (Figs.~1$a$,~$c$, and~$e$
correspond to the diffraction model),

$\sigma_{\Sigma}(q)=|F(q)|^{2}+\sigma_{2}(q)+\sigma_{3}(q)$ (Figs.~1$b$,~$d$, and~$f$ correspond
to the HEA model);
\setlength\parindent{4ex}

\vspace{5mm}
(v) the dashed curves plot the results of calculations based on the coupled-channels method with
a double folding potential, taken from~\cite{7} (Figs.~1$a$ and $b$) and~\cite{8} (Figs.~1$c$, $d$,
$e$, and $f$).

Table~1 displays values of the parameters of the diffraction model and the HEA model used in the
calculations of the curves plotted in Fig. 1.

It follows from a comparison of the calculated results with the experiments that both models give
a completely satisfactory description of the observed angular dependences of the cross sections:
they reproduce both the magnitude of $\sigma/\sigma_{R}$ and its shallow dependence on the scattering
angle. It can also be seen in Fig.~1 that the contribution of inelastic scattering is substantial only
for $\theta>(6\div8)^{\circ}$ and has no effect on the magnitude or position of the main maximum or the
first minimum. On the whole, taking inelastic scattering into account leads to a filling of the secondary
minima (smoothing of the oscillations) and some increase in the magnitude of the cross sections in the
indicated region.

\vspace{5mm}
\begin{figure}[!h]
\center
\includegraphics [scale=1.00] {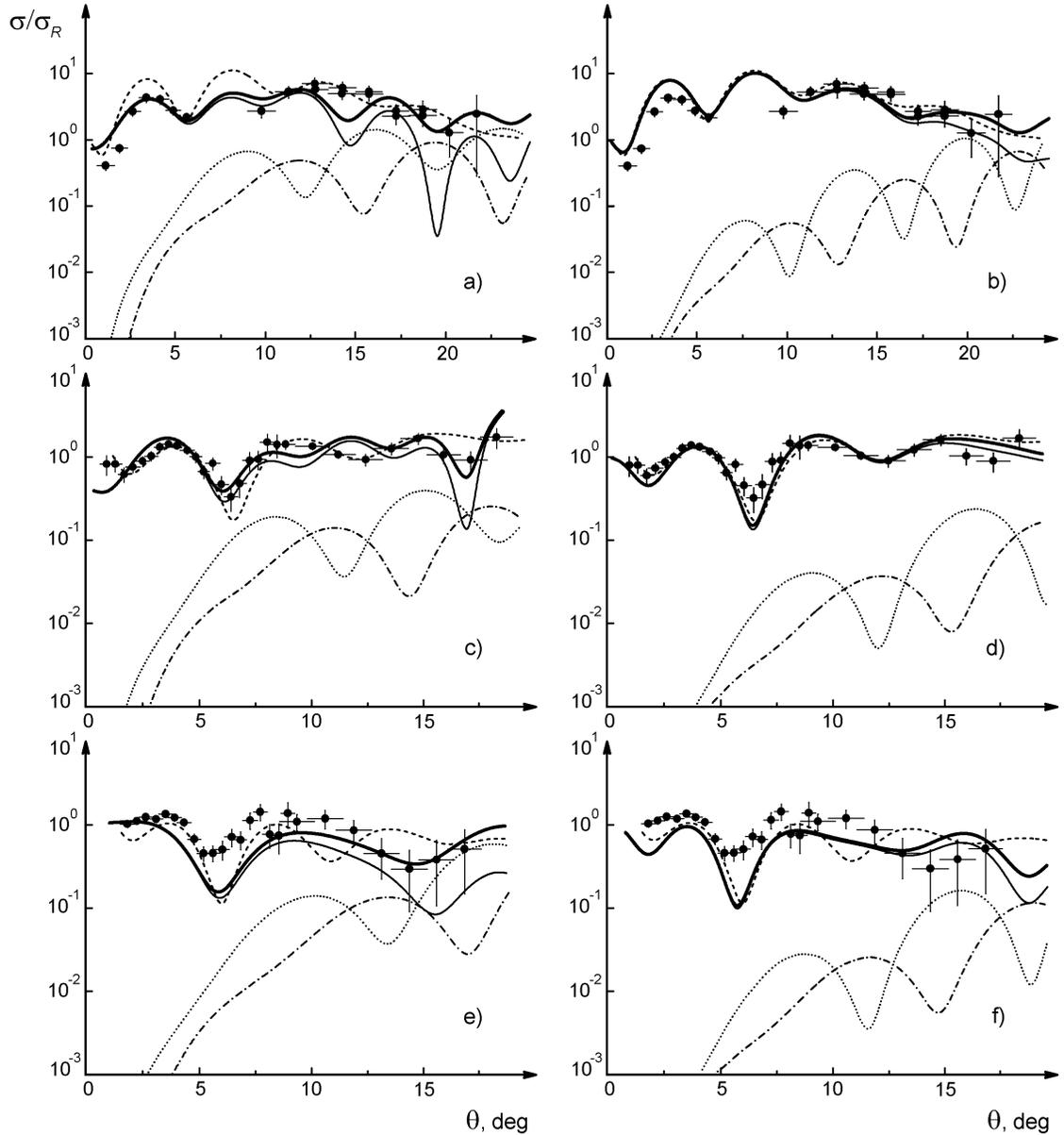}
\caption{Ratios of cross sections $\sigma/\sigma_{R}$ ($\sigma_{R}$ is the Rutherford cross section) for
scattering by $^{12}$\!C nuclei of $^{6}$\!He nuclei with energy \mbox{$T=494$~MeV} $(a,~b)$, of $^{7}$\!Be
nuclei with energy \mbox{$T=280$~MeV} $(c,~d)$, and of $^{8}$\!B nuclei with energy \mbox{$T=320$~MeV}
$(e,~f)$. Explanation of the types of curves is given in the text. The experimental data (points) were taken
from~\cite{7} ($^{6}$\!He) and~\cite{8} ($^{7}$\!Be, $^{8}$\!B).}
\label{fig1}
\end{figure}

\begin{table}[!h]
\caption{Values of the Model Parameters Used in the Calculations of the Cross Sections}
\vspace{-5mm}
\begin{center}
\begin{tabular}{|c|c|c|c|c|c|c|c|c|c|c|c|}
\hline
\hline & \multicolumn{5}{c|}{Diffraction model} & \multicolumn{4}{c|}{HEA model} \\
\cline{2-10}
\raisebox{1.5ex}[0cm][0cm]{Incident nucleus} & $r_0$,~fm & $\delta_1$ & $\gamma_1$ & $\delta_2$ & $\gamma_2$
& $N_{V}$ & $N_{W}$ & $r_W$,~fm & $a_W$,~fm \\
\hline
$^{6}$He & 0.8 & 0.7 & 0.49 & 0.080 & 0.072 & 1.0 & 1.0 & 0.91 & 0.61 \\
\hline
$^{7}$Be & 1.1 & 0.9 & 0.90 & 0.011 & 0.026 & 1.0 & 1.0 & 0.95 & 0.58 \\
\hline
$^{8}$B  & 0.8 & 0.9 & 0.90 & 0.220 & 0.044 & 1.0 & 1.0 & 0.89 & 0.75 \\
\hline
\hline
\end{tabular}
\end{center}
\label{tab1}
\end{table}

It has been stated more than once in the literature that the divergence of theory from
experiment for scattering angles $\theta<10^{\circ}$ remains an open question (see, for
example,~\cite{7} and the references therein). It can be seen in Figs.~1$a$ and $b$ that
the HEA model raises the values of the cross sections in the region of scattering angles
$\theta<10^{\circ}$. However, use of the diffraction approximation in the given case leads
to a satisfactory description of experiment due to the greater flexibility of the model --
for each of the clusters there are customized fitting parameters (see the table).

As for the curves presented in Fig.~1$e$, the divergence of the calculated results from
experiment at scattering angles $\theta<10^{\circ}$ is explained by the incompletely correct
model representation of the $^{8}$\!B nucleus as consisting of a $^{7}$\!Be core together
with one proton weakly bound to it. In our opinion, it would be more advantageous here to
use a three-cluster model of the $^{8}$\!B nucleus $(^{4}\text{He}+^{3}\text{He}+p)$~\cite{4}.

Thus, on the basis of the formalism developed in~\cite{4,5}, in the given work we have
successfully described the observed cross sections of quasi-elastic scattering of
$^{6}$\!He, $^{7}$\!Be, and $^{8}$\!B by $^{12}$\!C nuclei at intermediate energies of the
incident particles. The given approach can also be generalized to a three-cluster model of
the incident nucleus.


\end{document}